\documentclass[fleqn,usenatbib]{mnras}

\usepackage{newtxtext,newtxmath}

\renewcommand{\vec}[1]{\boldsymbol{#1}}

\usepackage[T1]{fontenc}

\DeclareRobustCommand{\VAN}[3]{#2}
\let\VANthebibliography\thebibliography
\def\thebibliography{\DeclareRobustCommand{\VAN}[3]{##3}\VANthebibliography}

\usepackage{graphicx}	
\usepackage{amsmath}	

\usepackage[usenames]{color}

\title[Near-core magnetic field strength in HD~43317]{Asteroseismic inference of the near-core magnetic field strength in the main sequence B star HD~43317}

\author[Lecoanet, Bowman \& Van Reeth]{Daniel Lecoanet$^{1,2,3}$\thanks{\href{mailto:daniel.lecoanet@northwestern.edu}{daniel.lecoanet@northwestern.edu}}, Dominic M. Bowman$^{4,3}$, Timothy Van Reeth$^{4,3}$
\\
$^{1}$Department of Engineering Sciences and Applied Mathematics, Northwestern University, Evanston IL 60208, USA \\
$^{2}$CIERA, Northwestern University, Evanston IL 60201, USA \\
$^{3}$Kavli Institute for Theoretical Physics, University of California, Santa Barbara, CA 93106, USA \\
$^{4}$Institute of Astronomy, KU Leuven, Celestijnenlaan 200D, B-3001 Leuven, Belgium \\
}

\date{Accepted XXX. Received YYY; in original form ZZZ}

\pubyear{2022}

\begin{document}
\label{firstpage}
\pagerange{\pageref{firstpage}--\pageref{lastpage}}
\maketitle

\begin{abstract}
About 10~per~cent of intermediate- and high-mass dwarf stars are observed to host a strong large-scale magnetic field at their surface, which is thought to be of fossil field origin. However, there are few inferences as to the magnetic field strength and geometry within the deep interiors of stars. Considering that massive stars harbour a convective core whilst on the main sequence, asteroseismology of gravity (g) modes is able to provide constraints on their core masses, yet it has so far not been used to probe the strength of their interior magnetic fields. Here we use asteroseismology to constrain an upper limit for the magnetic field strength in the near-core region of the pulsating and magnetic B~star HD~43317, based on the expected interaction of a magnetic field and its g~modes. We find a magnetic field strength of order $5 \times 10^5$~G is sufficient to suppress high-radial order g~modes and reproduce the observed frequency spectrum of HD~43317, which contains only high-frequency g~modes. This result is the first inference of the magnetic field strength inside a main-sequence star.
\end{abstract}




\section{Introduction}
\label{section: intro}

Magnetic fields are ubiquitous in astrophysics. However, observational constraints of internal magnetic fields of stars are limited. Large-scale magnetic fields have been detected at the surfaces of approximately 10\% of massive dwarf stars \citep{Power_MASTER, Shultz2012c, Shultz2018d, Wade2014a, Grunhut2017}, typically from high-resolution spectropolarimetry \citep{Wade2016a}. Most large-scale magnetic fields in massive stars have a predominantly dipolar geometry, which often is inclined to the rotation axis. The strengths of these fields are typically between 100~G and a few tens of kG. These large-scale magnetic fields likely extend deep within the radiative envelopes of massive stars, and are thought to be created during star formation \citep{Mestel1999a, Neiner2015d}, but also form during binary mergers \citep{Schneider_F_2019a}. Irrespective of their origin, magnetic fields have a large impact on stellar structure and evolution \citep{Maeder2005b, Keszthelyi2019}. 

Asteroseismology --- the study of stellar structure from oscillations --- is a valuable tool to ascertain the interior physics of stars across the Hertzsprung--Russell (HR) diagram \citep{ASTERO_BOOK, Aerts2021a}. Waves restored by buoyancy are known as Internal Gravity Waves (IGWs), and those that set up standing waves are known as gravity (g) modes, which are extremely sensitive to the Brunt-V{\"a}is{\"a}l{\"a} frequency profile. This makes g~modes excellent probes of the convective core masses of main-sequence massive stars \citep{Miglio2008a, Bowman2020c, Johnston2021b}. Furthermore, since massive stars are commonly moderate-to-rapid rotators, the strength of the Coriolis force can be quantified from its impact on g~modes \citep{Bouabid2013}. The asteroseismic diagnostic to constrain the rotation and core mass of main-sequence stars is the g-mode period spacing pattern, which typically spans up to a few dozen consecutive high-radial order prograde dipole g~modes in main-sequence stars. The `tilt' and asymptotic period spacing value probe the rotation rate and convective core mass, respectively \citep{VanReeth2016a, Papics2017a, Mombarg2019a, Pedersen2021a}. 

The strength and geometry of an interior magnetic field can leave an imprint on g~modes. \citet{Prat2019a, Prat2020a, VanBeeck2020a} have made theoretical predictions for frequency shifts due to a magnetic field in rotating main-sequence stars and predict them to be very small for high-frequency g~modes in such stars. So far, there have been no direct constraints of an internal magnetic field strength in the deep interior of a main-sequence star. In this work, we use the pulsating and magnetic B~star HD~43317 to make the first observational inference of an interior magnetic field strength in a main-sequence star using asteroseismology.


\section{HD~43317: a unique magnetic pulsator}
\label{section: HD43317}

HD~43317 is a B3.5\,V star with a precisely measured rotation period of $0.897673(4)$~d and a large-scale dipolar surface magnetic field with strength $B_{\rm p} = 1312 \pm 332$~G from dedicated spectropolarimetric observations \citep{Briquet2013, Buysschaert2017b}. High resolution spectroscopy of HD~43317 revealed it to likely be a single star with solar metallicity, a projected surface rotational velocity of $v\,\sin\,i = 115 \pm 9$~km\,s$^{-1}$, an effective temperature of $T_{\rm eff} = 17350 \pm 750$~K and a surface gravity of $\log\,g = 4.0 \pm 0.1$ \citep{Papics2012a}. HD~43317 was observed by the CoRoT mission \citep{Auvergne2009} which provided a 150.5-d light curve with an average cadence of 32~s. The light curve contained dozens of significant g-mode frequencies \citep{Papics2012a, Buysschaert2018c}. 

\citet{Buysschaert2018c} extracted all the significant g-mode frequencies from the CoRoT light curve of HD~43317. After delimiting the parameter space in the HR~diagram using the spectroscopic parameters, the authors performed forward asteroseismic modelling to ascertain a statistically best-fitting mass of $M = 5.8^{+0.1}_{-0.2}$M$_{\odot}$, core hydrogen content of $X_{\rm c} = 0.54^{+0.01}_{-0.02}$, and a parameterisation of convective boundary mixing (CBM) of $f_{\rm CBM} = 0.004^{+0.014}_{-0.002}$. Assumptions in their modelling strategy included: rigid rotation fixed at the measured surface rotation period (i.e. $\simeq0.62\Omega_{\rm crit,Roche}$); constant envelope mixing of $D_{\rm env} = 10$~cm$^{2}$\,s$^{-1}$; and a radiative temperature gradient within the CBM region. The best-fitting parameters were derived based on a large grid of non-rotating and non-magnetic {\sc MESA} structure models (r8118; \citealt{Paxton2011, Paxton2013, Paxton2015}) and corresponding theoretical pulsation mode frequencies using the stellar oscillation code {\sc GYRE} (v4.1; \citealt{Townsend2013b}) after identifying probable mode geometries of the detected g-mode frequencies. The effect of rotation was included within the {\sc GYRE} calculations using the Traditional Approximation for Rotation (TAR; \citealt{Townsend2013b}). We refer the reader to \citet{Buysschaert2018c} for full details.

HD~43317 is the only magnetic and rapidly rotating star pulsating in g~modes that has been modelled using asteroseismology \citep{Buysschaert2018c}. For (non-magnetic) pulsating stars in this mass range (i.e. SPB stars), much higher radial orders (i.e. $|n_{\rm pg}| \gg 1$) are more commonly observed \citep{Pedersen2021a}, making HD~43317 atypical given its distinct lack of high-radial order g~modes. 

\begin{figure}
    \centering
    \includegraphics{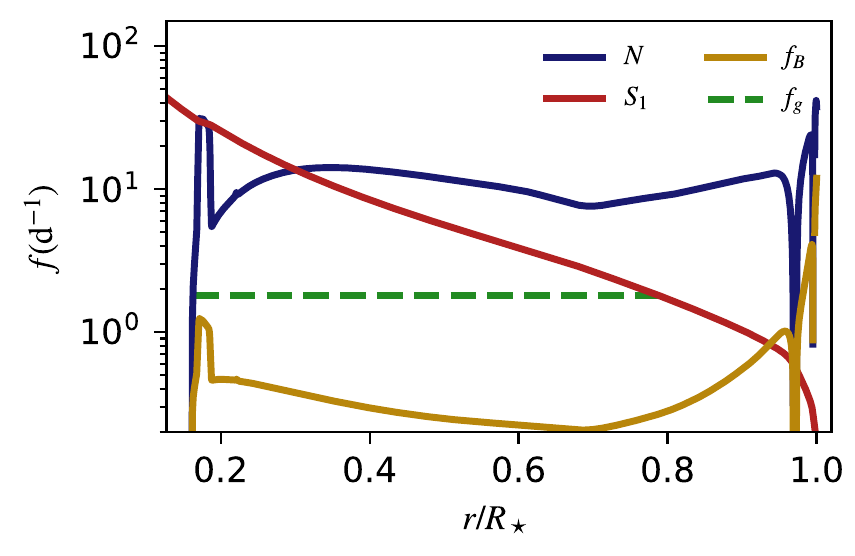}
    \caption{Propagation diagram for the magnetic SPB star HD43317 using the best-fitting asteroseismic model of \citet{Buysschaert2018c}. The green line shows the oscillation cavity of the lowest-frequency identified g~mode with frequency $f_g=0.692\, {\rm d}^{-1}$ in the inertial frame. The yellow line is an estimate of the magnetic interaction frequency for $\ell=1$, assuming the magnetic field is a dipole with strength $B_r=4.68\times 10^5 \, {\rm G}$ at $r=0.18R_\star$ (eqn.~\ref{eqn: f_B}).}
    \label{fig: propagation}
\end{figure}

In Fig.~\ref{fig: propagation}, we show the propagation diagram for the structure model determined by \citet{Buysschaert2018c} to best reproduce the observed g-mode frequencies and inferred mode geometry identification. In this paper we primarily study this best-fitting structure model and determine an upper limit of the magnetic field strength in the near-core region that satisfies the constraint that HD~43317 only has high-frequency g~modes. To test the robustness of our results, we also analyse an additional 19 models which satisfy the 2$\sigma$ confidence intervals on mass, age and CBM (see Table~2 of \citealt{Buysschaert2018c}). Although there are many free parameters that can be included when calculating a grid of structure models, we consider the asteroseismically calibrated models of \citet{Buysschaert2018c} to be sufficient to test the hypothesis that the interior magnetic field is strong enough to suppress low-frequency g~modes and explain why only high-frequency g~modes are observed.

\section{Interaction of internal gravity waves and magnetic fields}
\label{section: method}

The magnetic SPB star HD~43317 has only high-frequency g~modes observed in its amplitude spectrum of the CoRoT light curve, with radial orders spanning $-15 \leq n_{\rm pg} \leq -1$, which is atypical for SPB stars \citep{Buysschaert2018c}. The lowest-frequency g~mode is $f_g=1.806$~d$^{-1}$ in the corotating frame, which leads to the question of why lower frequency g~modes are not observed. Non-adiabatic pulsation calculations of HD~43317 using {\sc GYRE} \citep[v6.0.1;][]{Townsend2013b,Townsend2018,Goldstein2020} suggest a wider range of g~modes should be excited, with $n_{\rm pg}$ ranging from $-39 \pm 2$ to $-8 \pm 1$ for $(\ell,m)$ = $(2,-1)$ and from $-34 \pm 3$ to $-8 \pm 1$ for $(\ell,m)$ = $(1,-1)$.
We propose that these low-frequency g~modes are suppressed by a strong near-core magnetic field.

To understand the suppression of g~modes, we turn our attention to propagating IGWs. If an IGW can reflect off an upper and lower region of a star and its frequency satisfies a quantization condition, it can set up a standing mode, i.e., a g~mode.
\citet{Fuller2015} and \citet{Lecoanet2017} found that IGWs can have strong interactions with magnetic fields which prevent them from reflecting off the inner part of the star. This can suppress the dipole oscillation modes in RGB stars \citep[e.g.,][]{Stello2016}. They found that (ignoring rotation) if an IGW of frequency $f$ enters a region where
\begin{eqnarray}\label{eqn: f_B}
f \lesssim f_B = \frac{1}{2\pi} \sqrt{\frac{B_r}{\pi \rho} \frac{N\Lambda}{r}},
\end{eqnarray}
it can convert into a magnetic wave, preventing the formation of a standing g~mode.
Here $B_r$ is the radial magnetic field, $\rho$ is the density, $\Lambda=\sqrt{\ell(\ell+1)}$ with $\ell$ the spherical harmonic degree of the IGW, and $r$ is the local radius. This relation was used to estimate a magnetic field of $\sim 10^7 \, {\rm G}$ in the RGB star KIC~8561221 \citep{Garcia2014,Fuller2015}.
Assuming (as a rough approximation) the interior magnetic field of HD~43317 varies as $1/r^3$ like a dipole, Fig.~\ref{fig: propagation} shows that $f_B$ is greatest in the near-core region at $r\approx0.18R_\star$, if we only consider the oscillation cavities of the identified observed g~modes of HD~43317 which have frequencies above the green line. This indicates that the strongest interaction between IGWs and a dipolar magnetic field would occur in the near-core region where $N$ is large due to the chemical composition gradient.  In this region $f_B\sim 1 \, {\rm d}^{-1}$, indicating that magnetic and rotational effects are similar in strength. A similar calculation by \citet{Cantiello2016}, also found magnetic fields of $10^5\, {\rm G}$ can suppress modes with $f\lesssim 0.86 \, {\rm d}^{-1}$.

To refine this heuristic estimate, we calculate the linear waves of a rotating, magnetised star using the WKBJ approximation. We assume the background magnetic field is predominately dipolar, and take $B_r(r,\theta)=B_r(r)\cos(\theta)$. We assume all wave variables can be written as, e.g.,
\begin{eqnarray}\label{eqn: WKBJ}
U_r(r,\theta,\phi, t) = u_r(r,\theta)\,\exp\left[im\phi -i\omega t + i\epsilon^{-1}\int \psi(r) dr\right],
\end{eqnarray}
where $m$ is the azimuthal wavenumber, $\omega=2\pi f$ is the angular frequency, $u_r(r,\theta)$ is an amplitude, and $\psi$ is the phase.
The local radial wavenumber is $k_r=\epsilon^{-1}\psi'(r)$.
The WKBJ approximation is an exponentially-accurate asymptotic expansion in the small parameter $\epsilon\sim f/N$.
Assuming $N\sim B_r\sim \mathcal{O}(\epsilon^{-1})$ and all other parameters are $\mathcal{O}(1)$, the leading order oscillation equations are
\begin{align}\label{eqn: eigenvalue_first}
N(r)^2 u_r + k_r p &= 0, \\
 -i\omega\vec{u}_h+2\vec{\Omega}\vec{\times}\vec{u}_h +r^{-1}\vec{\nabla}_h p &= i B_r(r) \cos(\theta) k_r \vec{b}_h,\\
r^{-1}\vec{\nabla}_h\vec{\cdot}\vec{u}_h + i k_r u_r &= 0, \\
 -i\omega \vec{b}_h - i B_r(r)\cos(\theta) k_r \vec{u}_h &=- \eta k_r^2 \vec{b}_h.\label{eqn: eigenvalue_last}
\end{align}
Only the radial component of the magnetic field enters to leading order in $\epsilon$.

The perturbations are given by the radial velocity $u_r$, horizontal velocity and magnetic field $\vec{u}_h$ and $\vec{b}_h$, and pressure $p$. The rotation vector is $\vec{\Omega}$, $\vec{\nabla}_h$ gives the angular components of the gradient, and $\eta$ is the magnetic resistivity. For a given oscillation frequency $\omega$, the calculation is an eigenvalue problem at each radius with eigenvalue $k_r$. We solve this eigenvalue problem in spherical geometry with the {\sc Dedalus} code \citep{Vasil2019, Lecoanet2019, Burns2020}. The scripts used to generate the data and plots in this paper can be found at \url{https://github.com/lecoanet/HD43317-B}.


\section{Results}
\label{section: results}

\begin{figure}
    \centering
    \includegraphics{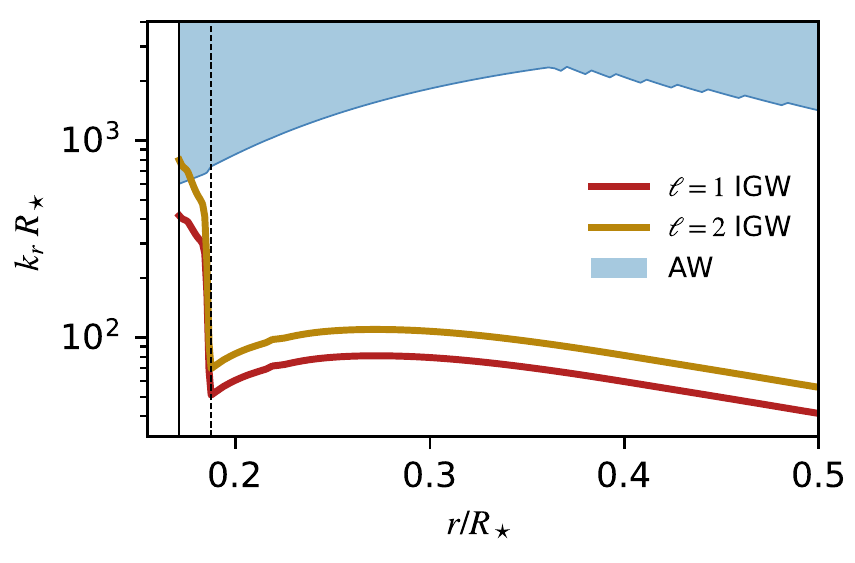}
    \caption{Radial wavenumber as a function of radius for $\ell=1$ and $\ell=2$ IGWs (red and yellow lines), and resonant Alfven waves (AWs; shaded blue region), assuming $B_r=4.68\times 10^5\, {\rm G}$ at $r=0.18R_\star$. The waves have $m=-1$ and $f=0.872 \, {\rm d}^{-1}$. The vertical solid and dashed lines are the radiative-convective boundary, and the boundary of the chemical composition gradient region. The IGWs do not interact with the AWs, so they can reflect off the radiative-convective interface and produce g~modes.}
    \label{fig: wavenumbers}
\end{figure}

In Fig.~\ref{fig: wavenumbers} we plot the radial wavenumber for different waves with $m=-1$ and $f=0.872$~d$^{-1}$ assuming a dipolar magnetic field with $B_r=4.68\times 10^5\, {\rm G}$ at $r=0.18R_\star$, and a magnetic resistivity $\eta=0$. We use the frequency of the observed g~mode with $\ell=2$ and $n_{\rm pg} = -15$; this is the lowest frequency $\ell=2$ mode observed by \citet{Buysschaert2018c} for HD~43317. The radial wavenumbers of the IGWs remain modest until they enter the chemical composition gradient region. At high wavenumbers, we find a continuum of resonant Alfven waves (AWs). These AWs have critical latitudes, and those with wavenumbers in the lower part of the blue shaded region of Fig.~\ref{fig: wavenumbers} have critical latitudes near the poles. When the radial wavenumber of an IGW enters the blue shaded region, it turns into a resonant AW, and its energy is spread out amongst the continuum modes. When this occurs, the wave cannot return to the surface as an IGW; thus the wave cannot set up a g~mode.
The $\ell=1$ and $\ell=2$ IGWs with $f=0.872$~d$^{-1}$ and a magnetic field strength of $B_r=4.68\times 10^5$~G do not turn into AWs. Thus, they can reflect off the radiative-convective interface, return to the surface as IGWs, and set up g~modes. Although the wavenumber of the $\ell=2$ IGW just barely enters the AW region, the IGW is equatorially confined, which means it has exponentially weak overlap with the near-polar resonant AWs. Finally, we note that the radial wavenumber changes abruptly at the edge of the chemical composition region (vertical dashed line in Fig.~\ref{fig: wavenumbers}). There may also be wave reflection off this point.

\begin{figure}
    \centering
    \includegraphics{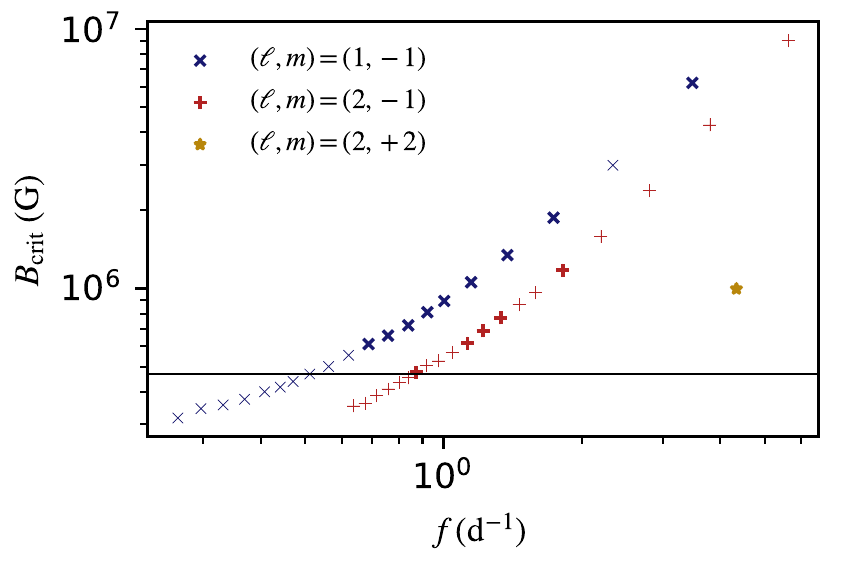}
    \caption{Critical magnetic field strength for different IGWs as a function of their frequency in the inertial frame. We study waves with the frequencies of the g~modes of HD~43317. Frequencies corresponding to observed (unobserved) modes are plotted with thick (thin) symbols. We infer a magnetic field with strength $B_r\approx4.7\times 10^5$~G at $r=0.18R_\star$ (horizontal line). A dipolar field of this strength would suppress g~modes below the horizontal line, explaining the lack of low-frequency modes.}
    \label{fig: critical}
\end{figure}

If the magnetic field strength is slightly higher, an IGW of the same frequency turns into an AW near the radiative-convective interface.
We define the critical magnetic field $B_{\rm crit}$ to be the threshold radial magnetic field strength at $r=0.18R_\star$, over which an IGW turns into an AW as it propagates into the star. This is a more precise version of the $f_B$ heuristic argument of Eqn.~\ref{eqn: f_B}. In Fig.~\ref{fig: critical} we plot $B_{\rm crit}$ for IGWs with frequencies which match theoretical mode frequencies calculated using GYRE. IGWs with frequencies of observed g~modes are plotted with thick symbols, and IGWs with frequencies of unobserved g~modes are plotted with thin symbols. The lowest frequency modes for each $\ell$ have the smallest $B_{\rm crit}$. For the $(\ell, m) = (2, -1)$ mode with frequency $f=0.872 \, {\rm d}^{-1}$, we have $B_{\rm crit}\approx 4.7\times 10^{5} \, {\rm G}$, whereas for the $(\ell, m) = (1, -1)$ mode with frequency $f=0.692 \, {\rm d}^{-1}$, we have $B_{\rm crit}\approx 6.1\times 10^{5} \, {\rm G}$.
This means, e.g., a near-core magnetic field of $\approx 4.7\times 10^5 \, {\rm G}$ would suppress all lower frequency $\ell=2$, $m=-1$ modes.
This gives a natural explanation for the lack of low-frequency g~modes in HD~43317.
The fact that the lowest frequency observed modes for each $\ell$ have similar $B_{\rm crit}$ supports this theory. In summary, the largest magnetic field consistent with the observed g~modes has a strength of $B_r\approx 4.7\times 10^5 \, {\rm G}$ at $r=0.18R_\star$.

There are two unobserved $\ell=1$ modes with $B_{\rm crit}>B_r$, which would not be suppressed by this magnetic field. These modes might not be observed because: (i) The modes are not excited to observable amplitudes; or (ii) The magnetic field near the radiative-convective interface is not entirely dipolar, leading to a different $\ell$-dependence of $B_{\rm crit}$. In addition, \citet{Buysschaert2018c} found three other oscillation modes which they could not identify with inertial-frame frequencies of $<0.692 \, {\rm d}^{-1}$. Our analysis suggests these are $m=-2$ modes which have large frequencies in the corotating frame, and thus are not strongly affected by the magnetic field.

\begin{figure}
    \centering
    \includegraphics{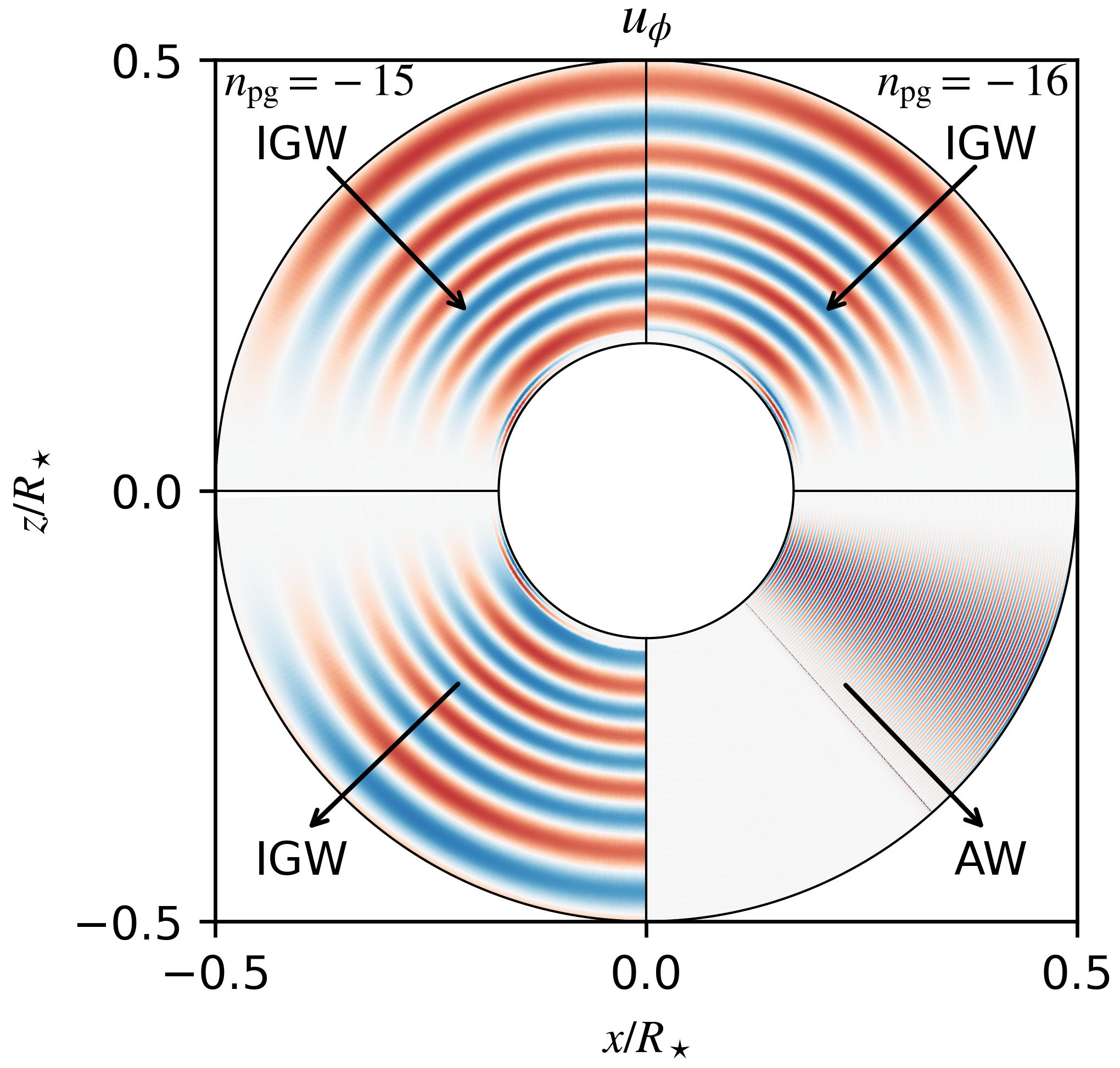}
    \caption{Azimuthal velocity structure for $m=-1$, $\ell=2$ waves with frequencies $f=0.872\, {\rm d}^{-1}$ (left; $n_{\rm pg}=-15$) and $f=0.840\, {\rm d}^{-1}$ (right; $n_{\rm pg}=-16$) in the inertial frame, assuming a dipole magnetic field with $B_r=4.68\times10^5 \, {\rm G}$ at $r=0.18R_\star$. The top panels show the incoming waves, and bottom panels show outgoing waves. The lower frequency wave, corresponding to the unobserved $n_{\rm pg}=-16$ mode, interacts strongly with the magnetic field when it enters into chemical composition gradient region, triggering a latitudinally-localised resonant AW. The AW is calculated only at a single radius near the radiative-convective interface, but is plotted over a range of radii to give a sense of its structure.}
    \label{fig: interior}
\end{figure}

To demonstrate the interaction between IGWs and a dipolar magnetic field, we plot two inward and outward propagating waves in Fig.~\ref{fig: interior}. On the left we plot $u_\phi(\phi=0)$ for an IGW with $f=0.872\, {\rm d}^{-1}$ in the inertial frame, which corresponds to the frequency of the $n_{\rm pg}=-15$ g~mode, the lowest-frequency observed $\ell=2$ mode. We plot the WKBJ solution described by Eqn.~\ref{eqn: WKBJ}, where we take $t=0$, and normalise $u_\phi(r,\theta)$ according to $\int |u_\phi|^2\sin(\theta) d\theta = 1$. The next-to-leading-order equations could be used to derive an equation for the amplitude as a function of $r$. The inward propagating wave (top left) has $k_r>0$, and is initialised with a phase of $\pi/2$ at $r=0.5$. The outward propagating wave (bottom left) has $k_r<0$, and matches the phase of the incoming wave at the radiative-convective interface. Although the wave has a small radial wavenumber in the chemical composition gradient region, it can reflect off the radiative-convective interface and set up a g~mode.

On the right of Fig.~\ref{fig: interior}, we plot $u_\phi$ in the same way for an IGW with $f=0.840\, {\rm d}^{-1}$, which corresponds to the frequency of the $n_{\rm pg}=-16$ g~mode, which is not observed. As this wave propagates inwards (upper right plot), it exhibits similar behaviour to the higher-frequency mode as the left hand side of the figure. However, near the radiative-convective interface, it undergoes a strong interaction with the magnetic field, turning into a resonant AW. In the bottom right quadrant of Fig.~\ref{fig: interior}, we show the structure of the resonant AW. We calculate the AW at a single radius near the radiative-convective interface, but plot it at a range of radii to illustrate its structure. To resolve the AW, we introduce a small magnetic resistivity of $\eta=3\times 10^{-10}$ into the problem. The AW has high radial wavenumber and a critical latitude near $45^{\circ}$. Because the resonant AWs form a continuum (Fig.~\ref{fig: wavenumbers}), we expect the AW to interact with many other AWs and spread across many latitudes. This means no IGW returns to the surface at this frequency, so there is no g~mode, in agreement with the observed properties of the star.

\begin{figure}
    \centering
    \includegraphics{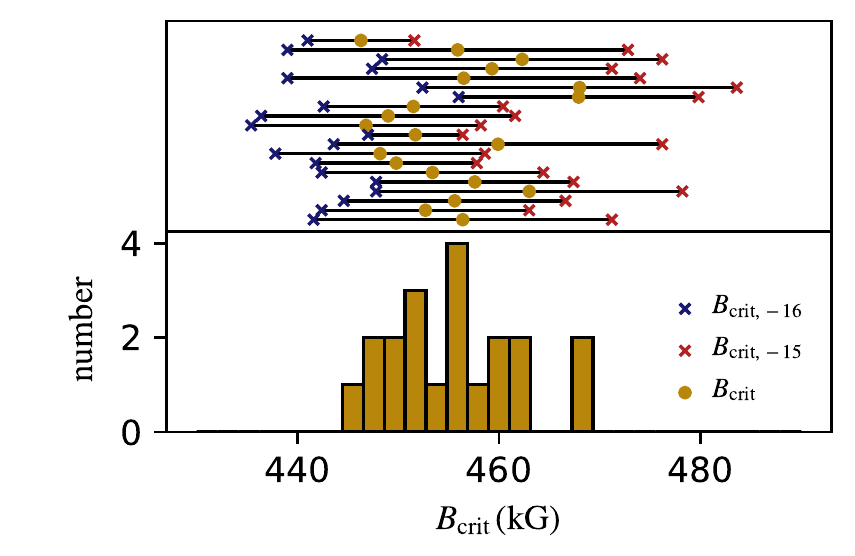}
    \caption{Top panel: Critical magnetic field strength for the 20 best fitting asteroseismic models. Blue (red) crosses show $B_{\rm crit}$ for IGWs with frequency of the $\ell=2$, $n_{\rm pg}=-16$ ($n_{\rm pg}=-15$) modes. The yellow circles show their mean, as estimate of the stellar magnetic field. Bottom panel: Histogram of the estimated magnetic field strength. We infer $B_r=456 \, {\rm kG}$ with an rms deviation across the 20 models of $6 \, {\rm kG}$.}
    \label{fig: histogram}
\end{figure}

There is uncertainty in the stellar structure of HD~43317 as inferred from asteroseismology.
Up to now, we have analysed the best-fitting structure model. To test the robustness of our inferred magnetic field strength, we calculate $B_{\rm crit}$ using the structure models from \citet{Buysschaert2018c} which satisfy the 2$\sigma$ confidence intervals on mass, age and CBM. In Fig.~\ref{fig: histogram}, we plot $B_{\rm crit}$ for the IGWs with the frequencies of the $(\ell, m)=(2,-1)$ modes with $n_{\rm pg}=-16$ and $n_{\rm pg}=-15$. Assuming the $n_{\rm pg}=-16$ mode is not observed at the surface due to the presence of a dipolar magnetic field, the magnetic field strength must be between these two $B_{\rm crit}$. To estimate this magnetic field strength, we plot the histogram of the mean of $B_{\rm crit}$ for these two modes. The average magnetic field strength across the 20 best-fitting models is $B_r=456 \, {\rm kG}$ in the near core region, $r\approx 0.18R_\star$. The root-mean-square deviation from this mean across the twenty models is $6 \, {\rm kG}$. The inferred magnetic field strength is insensitive to the different models.

\section{Conclusions}
\label{section: conclusions}

Magnetic massive stars are rare, and pulsating magnetic massive stars are rarer still. The pulsating, magnetic and rapidly rotating star HD~43317 provides a unique opportunity to infer the strength of the magnetic field in the near-core region based on the reported lack of observed low-frequency g~modes \citep{Buysschaert2018c}. To reproduce the lowest-frequency observed g~mode based on the previous forward asteroseismic modelling, we find a critical magnetic field strength of approximately $5 \times 10^{5}$~G is needed in the near-core region for HD~43317. This inference is the first of its kind for a main-sequence star and serves as a valuable proof-of-concept for two reasons. First, pulsation modes can diagnose interior magnetic fields. Second, the interaction of waves with a strong internal magnetic field can explain why only certain radial order modes are visible at the surface of HD~43317. 

What is the physical origin of this magnetic field? HD~43317 has a surface magnetic field of strength $\approx 1.3 \, {\rm kG}$. If the strength of the field scales as $\sim r^{-3}$ from the surface to the interior of the star, the field in the near-core region would be $B_r\sim 200 \, {\rm kG}$, which is smaller than our inferred field strength by a factor of $\approx 2$. It is also possible that the near-core magnetic field is enhanced by a core convective dynamo. 3D simulations of B stars find magnetic field strengths of $\approx 200 \, {\rm kG}$ \citep{Augustson2016}. However, \citet{Featherstone2009} found the presence of a fossil field can enhance the dynamo magnetic field up to $\approx 500 \, {\rm kG}$, as we find here. Our results support the conclusion that HD~43317's near-core magnetic field is a stronger-than-normal core convective dynamo induced by a strong fossil field, which is observed at the star's surface.

In the future, we anticipate that this methodology can be expanded and applied to potentially many more pulsating magnetic stars being discovered by the ongoing NASA TESS mission \citep{Ricker2015}. Thus, measuring near-core field magnetic field strengths sampling mass and age on the upper-main sequence is now within reach.


\section*{Acknowledgements}

The authors thank Jim Fuller, Geoff Vasil, Kyle Augustson, Matteo Cantiello, Stephane Mathis, and Conny Aerts for useful discussions regarding IGWs, magnetic fields, and dynamo physics. DL is supported in part by NASA HTMS grant 80NSSC20K1280. DMB and TVR gratefully acknowledge funding from the Research Foundation Flanders (FWO) by means of senior and junior postdoctoral fellowships with grant agreements 1286521N and 12ZB620N, respectively, and FWO long stay travel grants V411621N and V414021N, respectively. Computations were conducted with support by the NASA High End Computing (HEC) Program through the NASA Advanced Supercomputing (NAS) Division at Ames Research Center on Pleiades with allocation GIDs s2276. The authors are grateful to the staff and scientists at the Kavli Institute for Theoretical Physics, University of California, Santa Barbara, and to the (other) organisers of the `Probes of Transport in Stars' program, which hosted them during this work. This research was supported in part by the National Science Foundation under Grant No. NSF PHY-1748958.

\section*{Data Availability}

All generated data and scripts used to make the figures for this paper are provided at \url{https://github.com/lecoanet/HD43317-B}.



\bibliographystyle{mnras}
\bibliography{Bfield_inference} 






\bsp	
\label{lastpage}
\end{document}